\documentclass[a4paper,11pt]{article}
\usepackage{jheppub} 
\usepackage{lineno}
\usepackage{feynmf}
\usepackage{subcaption}
\usepackage{caption}

\arxivnumber{2311.12797} 

\title{\boldmath Tilted Dirac superconductor at quantum criticality: Restoration of Lorentz symmetry}

\author[a]{Pablo Reiser}
\author[a,b,1]{Vladimir Juri\v{c}i\'c \note{Corresponding author.} }

\affiliation[a]{Departamento de F\'isica, Universidad T\'ecnica Federico Santa Mar\'ia, Casilla 110, Valpara\'iso, Chile.}

\affiliation[b]{Nordita, KTH Royal Institute of Technology and Stockholm University,
Hannes Alfvéns väg 12, SE-106 91 Stockholm, Sweden}

\emailAdd{pablo.reiser@sansano.usm.cl}
\emailAdd{vladimir.juricic@usm.cl}

\abstract{Lorentz symmetry appears as a quite robust feature of the strongly interacting Dirac materials even though the lattice interactions break such a symmetry.  We here demonstrate that the Lorentz symmetry is restored at the quantum-critical point (QCP) separating the tilted Dirac semimetal, breaking this symmetry already at the noninteracting level, from a gapped $s-$wave superconducting instability. To this end, we employ a one-loop $\epsilon=(3-D)-$expansion close to  the $D=3$ upper critical dimension of the corresponding Gross-Neveu-Yukawa field theory.  In particular, we show that the tilt parameter is irrelevant and ultimately vanishes at the QCP separating the two phases. In fact, as we argue here, such a Lorentz symmetry restoration may be generic for the strongly interacting tilted Dirac semimetals, irrespective of whether they 
feature mirror-symmetric or mirror-asymmetric tilting, and is also insensitive to whether the instability represents an insulator or a gapped superconductor. The proposed scenario can be tested in the   quantum Monte Carlo simulations of the interacting  tilted Dirac fermion lattice models. 
 }

\begin{document}
\maketitle
\flushbottom

\section{Introduction}
\label{sec:intro}

The Dirac crystals featuring Dirac quasiparticles as their low-energy excitations  exhibiting a linear  dispersion  opened up a new paradigm in the modern condensed matter physics~\cite{CastroNeto_09,Balatsky2014,armitage2018}. They  also bridge the gap between the seemingly unrelated phenomena in the realms of high-energy and condensed-matter physics through the emergence of the Dirac and Weyl quasiparticles, being the prime examples in this respect. Such Dirac materials feature an emergent relativistic-like Lorentz symmetry because of the Dirac quasiparticles' linear energy-momentum dispersion, $E({\bf k})=v_F k$, with the Fermi velocity, $v_F$, typically a few hundred times smaller than the speed of light, playing the role of an effective  velocity of light in their crystalline universe. Such a relativistic-like Lorentz symmetry turns out to be quite a robust feature of the Dirac materials even when the electron-electron interactions that explicitly break it are taken into account. For instance, the  Lorentz symmetry emerges in the deep infrared regime for the two-dimensional (2D)  Dirac fermions coupled via the long-range Coulomb interaction, where the speed of light takes over as the common velocity for both the Dirac fermions and the photons~\cite{Gonzalez1994}. Furthermore, for Yukawa (short-range) interacting Dirac  fermions,  the quantum-critical point (QCP) separating  a semimetallic and a strongly-coupled gapped (insulating or superconducting) phase is Lorentz symmetric~\cite{SSLee2007,Isobe2012,Roy2016,Roy2018,Roy2020}. This issue has been also recently addressed beyond the realm of the usual Hermitian Dirac materials, by taking into account the effects of the non-Hermiticity, that may arise, e.g.  from the coupling with an environment, resulting in the Lorentz symmetry emerging  at low energies~\cite{juricic2023yukawalorentz,murshed2023quantum}. These results point towards a ubiquitous emergence of the Lorentz symmetry at low energy in the Dirac materials, even though such a symmetry is absent at the lattice (UV) energy scale.  

To obtain further insights into the ubiquity of such an emergent Lorentz  symmetry, we here consider tilted Dirac materials, featuring low-energy quasiparticles with tilted Dirac dispersion, explicitly breaking this symmetry at the lattice scale already at the noninteracting level. Such an explicit symmetry breaking is realized by the direction of the tilt term in the dispersion of the tilted Dirac fermions (TDFs), which is dictated by the underlying crystalline lattice, and is represented by a single term $\sim\alpha$, with $\alpha$ as the tilt parameter, in a Dirac Hamiltonian, as shown in Eqs.~\eqref{hamiltonian} and \eqref{eq:hamiltonian-symmbreaking}. Introduction of such a tilt term may be considered as possibly the simplest Lorentz-symmetry-breaking deformation of an otherwise  Lorentz-symmetric noninteracting  Dirac Hamiltonian. Its effects have been studied in various contexts, for instance, it was argued that TDFs may represent a platform for the simulation of the curved space-time   in  condensed-matter systems~\cite{Volovik2016,Nissinen2017,Jafari-PRB2019,Ojanen-PRR2019,Jafari-PRR2020,Schmidt-SciPost2021,Volovik2021,Meng-PRR2022,vanWezel-PRR2022,vanderBrink-PRB2023}, while the quantum transport~\cite{Bergholtz-PRB2015,Carbotte-PRB2016,Stoof-PRB2017,Rodriguez-Lopez2020,Maytorena-PRB2021,Hao-Ran-PRB2022,Portnoi-PRB2022},  the effects of the long- and short-range components of the Coulomb interaction~\cite{Fritz-PRB2019,Rostami-PRR2020,RudneiRamos-PRB2021,RudneiRamos-PRB2023}, charged impurities~\cite{Bergholtz-PRB2017,Fritz-PRB2017,GZLiu-PRB2018,YWLee-PRB2019}, and the magnetic field~\cite{Goerbig-PRB2008,Carbotte-PRB-2018,Nag_2021} have also been addressed. 
We emphasize that the tilted Dirac dispersion has been theoretically   investigated in the materials using first-principle techniques~\cite{Soluyanov2015,Schnyder2017,Geilhufe-2018}.  In fact, the TDFs have been realized in a plethora of quantum materials~\cite{Koji-2006,Akito-2007,Koji-2014,Deng2016,Yun-Wu2016,ChenluWang-2016,Kenan-Zhang2017,Bihlmayer-PRB2017}, both with the subcritical tilt (type-1), featuring the point-like Fermi surfaces, and in overtilted materials  (type-2), with the electron- and hole-like Fermi surfaces,  putting the tilted Dirac crystals within the landscape of the quantum materials. 

In this article, we explore the effects of the  $s-$wave superconducting instability in the type-1 tilted Dirac materials. Our main result is that at the strong-coupling QCP separating the tilted Dirac semimetal and the $s-$wave SC phase, the tilt operator becomes  \emph{irrelevant} and therefore the Lorentz symmetry gets restored. To address this problem, we employ the Gross-Neveu-Yukawa (GNY) theory~\cite{Zinn_book,Rosenstein1993,sslee-2007,RJH-PRB2013,RY2013,Ponte_2014,Jian2015,Zerf2016}, since  this instability necessarily occurs at strong coupling, as can be seen already at the mean-field level [Eq.~\eqref{eq:MF-OP}], due to a vanishing density of states (DOS) at zero energy [Eq.~\eqref{eq:DOS}]. To account for the effects of both fermionic and order-parameter (OP) fluctuations in the vicinity of such a QCP,  since both the Yukawa and  $\Phi^4-$couplings [Eq.~\eqref{eq:renormalized-action}] are marginal  in three spatial dimensions ($D=3$), we use the renormalization group (RG) analysis within the $\epsilon=(3-D)$-expansion. We demonstrate that the tilt parameter in the   mirror-symmetric [Eq.~\eqref{hamiltonian}] and mirror-symmetry-breaking [Eq.~\eqref{eq:hamiltonian-symmbreaking}] tilted Dirac semimetals are   irrelevant at the superconducting  QCP, with the corresponding RG $\beta-$function given by Eq.~\eqref{eq:beta-alpha} [see also Fig.~\ref{Fig:Kalpha}], and the flow shown in Fig.~\ref{fig:RGflow}, therefore implying the restoration of the Lorentz symmetry at such a QCP. To further corroborate the ubiquity of this  scenario, we show that  the transition into a CDW phase also features the emergent Lorentz symmetry at low energy, albeit with a slower rate of the restoration, as shown in Eq.~\eqref{eq:beta-alpha-CDW}, see also Fig.~\ref{fig:RGflowCDW}  and Fig.~\ref{Fig:Kalpha}. We argue that such an emergence  of the Lorentz symmetry  is generic for the TDFs interacting via a short-range Coulomb interaction, with the rate that depends on whether   the OP commutes or anticommutes with the tilt term. Finally, the anticommutation makes  the rate of the restoration faster, as can be seen explicitly in the case of the SC and CDW instabilities.   

The paper is organized as follows. In Sec.~\ref{sec:low-energy-theory}, we introduce the low-energy theory for the 2D TDFs, while in Sec.~\ref{sec:mean-field}, we analyze the short-range interactions at the mean-field level. We then proceed with the setup of the quantum-critical GNY field theory in Sec.~\ref{sec:GNY}, and in Sec.~\ref{sec:RG analysis} we perform its RG analysis. Our conclusions are presented in Sec.~\ref{sec:conclusions}, and additional technical details are relegated to the Appendix~\ref{A}.

\section{Tilted Dirac superconductor: The low-energy theory}~\label{sec:low-energy-theory}

We consider a low-energy effective theory for the Dirac fermion quasiparticles around the two inequivalent valleys, as for instance,  two Dirac points at the two inequivalent corners of the Brillouin zone  in graphene.  To construct the Nambu-doubled basis for the Dirac superconductor, we first write  an eight-component Dirac spinor given by~\cite{Szabo-RoyPRB2021}
\begin{equation}
    \Psi_{\omega,\mathbf{k}}=\begin{pmatrix}
        c^{(+)}_{\uparrow} & c^{(-)}_{\uparrow} & c^{(+)}_{\downarrow} & c^{(-)}_{\downarrow}
    \end{pmatrix}^{T},\quad c^{v}_{s}=\begin{pmatrix} 
    a^{v}_{s}, & b^{v}_s
    \end{pmatrix},
\end{equation}
where \( a^{v}_s \) and \(b^{v}_s\) are the fermionic  fields corresponding to two different ($A$ and $B$) sublattices with valley index \(v=(\pm)\), and spin projection \(s=\uparrow,\downarrow\). Then, we introduce a particle-hole degree of freedom and write a doubled, sixteen-component Dirac-Nambu spinor as
\begin{equation}\label{nambu}
\Psi_{\omega,\mathbf{k}}\rightarrow\Psi_{\text{Nam}}=\begin{pmatrix}
        \Psi_{\omega,\mathbf{k}}\\
        \Gamma_{210}\Psi^{*}_{-\omega,-\mathbf{k}}
    \end{pmatrix}.
\end{equation} 
We  conveniently write matrices acting on the sixteen-dimensional Dirac-Nambu spinors as \(\Gamma_{\mu\nu\lambda\rho}=\eta_{\mu}\otimes \sigma_{\nu}\otimes \tau_{\lambda}\otimes \alpha_{\rho}\). Here,  \(\{\eta\}\), \(\{\sigma\}\), \(\{\tau\}\), and \(\{\alpha\}\)  are the sets of Pauli matrices operating on particle-hole, spin, valley and sublattice degrees of freedom, respectively, with \(\mu,\nu,\lambda,\rho=0,1,2,3\), and the index \(0\) corresponding  to the \(2\times2\) unity matrix. Notice that the lower component of the Dirac-Nambu spinor in Eq.~\eqref{nambu} is a time-reversal partner of the upper one, since the matrix \(\Gamma_{210}=\sigma_{2}\otimes \tau_{1} \otimes \alpha_0\) is the unitary part of the time reversal operator in the undoubled basis. Furthermore, the form of the $\Gamma-$matrices dictates that any two of them either commute or anticommute. To simplify notation, we suppress subscript ``Nam" hereafter,  \(\Psi\equiv\Psi_{\text{Nam}}\) and \(\Psi^{\dagger}\equiv\Psi_{\text{Nam}}^{\dagger}\). 

To make a more direct connection to the relativistic Dirac field theory, we notice that in this way defined $16(=2\times8)$-component  Euclidean Dirac fermions in  one imaginary-time and   two spatial dimensions   correspond to 8 copies of two-component spin-$1/2$ relativistic Dirac fermions living in the $(2+1)-$dimensional Minkowski space, with the sublattice degrees of freedom directly translating  into the (real) spin degrees of freedom of the relativistic fermions~\cite{Semenoff1984}, therefore justifying the label "pseudo-spin" for the sublattice index. Explicitly, the matrices $\alpha$ act on the Lorentz spinor indices provided this identification, with their relativistic version being $\tilde{\alpha}^1=-i\alpha_2$, $\tilde{\alpha}^2=i\alpha_1$, while $\tilde{\alpha}^0=\alpha_3$. They  obey the  Clifford algebra, $\{\tilde{\alpha}^\mu,\tilde{\alpha}^\nu\}=2\eta^{\mu\nu}$,  with respect to the  metric $\eta_{\mu\nu}={\rm diag}(+,-,-)$ in the $(2+1)-$dimensional Minkowski spacetime.  
On the other hand, in the $(3+1)-$dimensional Minkowski spacetime, a minimal representation of the $\Gamma-$matrices is four-dimensional, with  their explicit form that can be  chosen to be in the  so-called chiral representation, $\tilde{\Gamma}^\mu\equiv(\tilde{\Gamma}^0,\tilde{\Gamma}^j)$, $j=1,2,3$,  with $\tilde{\Gamma}^0=\tau_2\otimes\alpha_0$, ${\tilde\Gamma}^j=i\tau_1\otimes\alpha_j$, and the $\tau$ ($\alpha$) matrices acting on the chiral (spin) subspaces. This representation is especially important as it permits to establish an explicit connection between  the degrees of freedom in the relativistic Dirac theory and  the  low-energy (Euclidean) effective field theories   of Dirac and Weyl semimetals~\cite{armitage2018}, see also the discussion after Eq.~\eqref{eq:renormalized-action}.      

Given the  Dirac-Nambu representation in Eq.~\eqref{nambu}, we can write the noninteracting low energy effective Hamiltonian for the tilted Dirac fermions in the form \( \hat{H}=\sum_{\mathbf{k}}\Psi^{\dagger}(\mathcal{\hat{H}}(\mathbf{k})-\mu_{\rm tot}\Gamma_{3000})\Psi\), with  \(\mu_{\rm tot}\) as the chemical potential, and 
\begin{equation}\label{hamiltonian}
    \mathcal{\hat{H}}(\mathbf{k})=v\Gamma_{3031}k_x + v\Gamma_{3002}k_y +\alpha v\Gamma_{3030}k_x,
\end{equation}
where \(v\) is the Fermi velocity and  \(\alpha\) represents the (dimensionless) tilt parameter. We here consider only  case with \(|\alpha|<1\) (type-1) featuring the point-like Fermi surfaces with the TDF quasiparticle excitations, while  the overtilted (type-2) system with $|\alpha|>1$ hosts  a particle- and a hole-like Fermi surfaces. Notice that the tilt term can be thought of as a momentum-dependent chiral chemical potential, and it commutes with the (total) chemical potential represented by the matrix $\Gamma_{3000}$. We hereafter set $v=1$ and consider the system at half filling, $\mu_{\rm tot}=0$, unless otherwise stated. 

The form of the  single-particle Hamiltonian implies its invariance under time reversal, generated by $\mathcal{T}=\Gamma_{0210}{\rm K}$, with ${\rm K}$ as the complex conjugation, SU(2) spin rotations generated by \(\Gamma_{0s00}\), with \(s=\{1,2,3\}\), as well as the valley (sublattice) exchange, represented  by $V=\Gamma_{0010}$, $k_x\rightarrow-k_x$ ($S=\Gamma_{0001}$, $k_y\rightarrow-k_y$)~\cite{HJR-PRB2009}. Finally, the unitary particle-hole symmetry is represented by $\mathcal{C}=\Gamma_{1000}$, so that \(\{\mathcal{C}, \mathcal{\hat{H}}({\bf k})\}=0\), implying that  its anti-unitary counterpart is represented by the operator $\mathcal{C}\mathcal{T}$. Notice that   $\tilde{\mathcal{C}}=\Gamma_{2000}$ can also be taken as a unitary particle-hole operator, which is a consequence of the U(1) symmetry of the noninteracting Hamiltonian generated by the matrix $\Gamma_{3000}$ representing the electric charge operator in the Dirac-Nambu representation employed here. Finally, we emphasize that the choice of the tilt term ($\sim\alpha$) in Eq.~\eqref{hamiltonian} ensures its invariance under all the mentioned symmetries.


We point out that the mirror-symmetry breaking tilted Dirac Hamiltonian, with the sign of the tilt equal for all the valleys (or other fermion flavors) is obtained by breaking the valley exchange or mirror symmetry discussed above, and we refer to it as \emph{asymmetric} tilt. The corresponding Hamiltonian then takes the form 
\begin{equation}\label{eq:hamiltonian-symmbreaking}
    \mathcal{\hat{H}_{\rm AS}}(\mathbf{k})=v\Gamma_{3031}k_x + v\Gamma_{3002}k_y +\alpha v\Gamma_{3000}k_x,
\end{equation}
with the tilt term proportional to the matrix representing  the chemical potential. Therefore, the asymmetric  tilt term  effectively represents a momentum-dependent chemical potential. As we show in the following, our conclusions regarding the restoration of the Lorentz symmetry in the case of the interacting  fully symmetric tilted Dirac Hamiltonian [Eq.~\eqref{hamiltonian}] equally apply to the asymmetrically tilted Hamiltonian in Eq.~\eqref{eq:hamiltonian-symmbreaking}, as follows from the form of the corresponding propagator, Eq.~\eqref{fermion-prop-asymm}. 

\section{Mean-field analysis}~\label{sec:mean-field}

To set the stage for the analysis of the strong-coupling superconducting instability of the tilted Dirac semimetal, we first discuss  mean-field (MF) theory  for a transition from the semimetallic   to a fully gapped ordered state,  which is independent of the nature of the state, i.e. whether the state is an insulator or a superconductor. For concreteness,  we here subscribe to the $s-$wave superconducting order parameter, which reads 
\begin{equation}
\phi(\mathbf{r},\tau)=\left\langle{\Psi^{\dagger}(\mathbf{r},\tau)(\Gamma_{1000}\cos{\varphi}+i\Gamma_{2000}\sin{\varphi})\Psi(\mathbf{r},\tau)}\right\rangle,
\end{equation}
where \(\varphi\) is the U(1) phase of the superconducting OP. This OP anticommutes with the particle number operator \(N=\Gamma_{3000}\),  and commutes with all the three generators of spin rotations, \(\Gamma_{0s00}\), as dictated by its s-wave nature. Notice that both matrices in the $s$-wave OP \emph{anticommute} with the tilted Dirac Hamiltonian in Eq.~\eqref{hamiltonian} which ensures that the underlying ordered state is fully gapped. 
In the MF analysis, we set the OP to be a constant, $\phi(\mathbf{r},\tau)\equiv m$, and fix the SC phase $\varphi=0$. We assume  that such a condensation is propelled by an effective 
local density-density  interaction of the form
\begin{equation}\label{Sinteraction}
\mathcal{H}_{\text{int}}=\tilde{g}\int d^2 \mathbf{r}\left[\Psi^{\dagger}(\mathbf{r},\tau)\,\Gamma_{1000}\,\Psi(\mathbf{r},\tau)\right]^2. 
\end{equation}
Then, we formulate the MF theory for the transition into such a state  by taking  the Hamiltonian  
\begin{equation}
{H}_{\rm MF}=\int d^2{\bf r}\,\Psi^{\dagger}(\mathbf{r})\,\Big\{ \mathcal{\hat{H}}(\mathbf{k}\rightarrow -i\nabla_{{\bf r}}) \Big\}\Psi(\mathbf{r}) + \mathcal{H}_{\text{int}},
\end{equation}
with the first term representing the free tilted Dirac Hamiltonian in Eq.~\eqref{hamiltonian}, together with the short-range interaction part, given by Eq.~\eqref{Sinteraction}. We then perform a Hubbard-Stratonovich transformation in the $s-$wave channel to obtain the corresponding energy of the ground state at half filling ($\mu=0$) in the form, 
\begin{align}
    \mathcal{E}[m]&=-\frac{1}{2\tilde{g}}m^2+2 \sum_{w=\pm}\int \frac{d^2 \mathbf{k}}{(2\pi)^2}\left\{w \alpha k_x+\sqrt{k^2 +m^2}\right\}\nonumber\\
    &= -\frac{1}{2\tilde{g}}m^2+4\int \frac{d^2 \mathbf{k}}{(2\pi)^2}\sqrt{k^2 +m^2}.
\end{align}
Therefore,  the MF ground state energy is tilt-independent, yielding the same form of the gap equation as in the untilted case, after minimization with respect to $m$, and implies that
\begin{equation}\label{eq:MF-OP}
m\sim (\tilde{g}-\tilde{g}_c),
\end{equation}
with the tilt-indenpendent  critical coupling $\tilde{g}_c\sim\Lambda$ scaling with the ultraviolet cutoff, defining  the energy scale up to which the Dirac dispersion holds. Therefore, the transition into the  $s-$wave (and, more generally, any gapped) phase takes place at a  strong coupling, 
which is a consequence of the vanishing  density of states (DOS) for the tilted Dirac fermions
\begin{equation}~\label{eq:DOS}
    \rho(E,\alpha)=-\frac{1}{\pi}\lim_{\delta \to 0}\Im {\text{Tr}\,G_F(i\omega\rightarrow \omega + i\delta,\mathbf{k})}=\frac{4}{\pi}\frac{|E|}{(1-\alpha^2)^{3/2}},
\end{equation}
for a single 16-component Dirac flavor. 
It is therefore expected that at any finite-tilt fixed point of the RG, the critical value of the interaction decreases with respect to the untilted ($\alpha=0$) value. Indeed  this is the case for unstable fixed points at finite tilt corresponding to  the $s-$wave and the CDW OPs, which can be obtained from Eq.~\eqref{eq:beta-g2} and \eqref{eq:g2-cdw}, respectively, which are, however, rendered unstable, as the tilt is irrelevant for both the $s-$wave and the CDW OPs, see Eqs.~\eqref{eq:beta-alpha} and~\eqref{eq:beta-alpha-CDW}.


\section{Gross-Neveu-Yukawa quantum-critical field theory}~\label{sec:GNY}

We now address the behavior of the \((2+1)-\)dimensional  tilted  $s-$wave Dirac superconductor at the quantum-critical point within the framework of the Euclidean (imaginary-time) Gross-Neveu-Yukawa (GNY) field theory.

\subsection{  $S-$wave superconducting instability }

First, since both noninteracting Hamiltonian and the OP are spin singlets,  we  discard the spin subspace hereafter.  We therefore consider  eight-component massless Dirac fermions  endowed with \(N_F\) copies, which interact with a complex bosonic order parameter field via Yukawa interaction ($g$). The $s-$wave superconducting OP is in the form given by Eq.~\eqref{eq:MF-OP}, which after discarding the spin subspace, reads 
\begin{equation}
\phi(\mathbf{r},\tau)=\left\langle{\Psi^{\dagger}(\mathbf{r},\tau)(\Gamma_{100}\cos{\varphi}+i\Gamma_{200}\sin{\varphi})\Psi(\mathbf{r},\tau)}\right\rangle.
\end{equation}
Hereafter, \(\Psi\equiv \Psi(\mathbf{r},\tau)\), \(\Psi^{\dagger}\equiv\Psi^{\dagger}(\mathbf{r},\tau)\) and \(\phi\equiv\phi(\mathbf{r},\tau)\). The dynamics of the system is then described by an imaginary time (Euclidean) action
\begin{equation}
    S=S_{F}+S_{Y}+S_{B},
\end{equation}
with the noninteracting fermionic part of the form 
\begin{equation}\label{Sfermion}
    S_{F}=\int d\tau d^{D} \mathbf{r} \,\,\Psi^{\dagger}\left[\partial_{\tau}+\mathcal{\hat{H}}(\mathbf{k}\rightarrow -i\nabla_{\mathbf{r}})\right]\Psi,
\end{equation}
where \(\tau\) is the imaginary time and \(\mathcal{\hat{H}}(\mathbf{k})\) represents the Hamiltonian for noninteracting massless TDFs in Eq.~\eqref{hamiltonian}. The coupling of the massless Dirac fermions to the OP fluctuations takes the characteristic Yukawa form,  
\begin{equation}\label{eq:Yukawa}
    S_{Y}=g\int d\tau d^{D}\mathbf{r}[(\text{Re}\phi)\Psi^{\dagger}\Gamma_{100}\Psi+(\text{Im}\phi)\Psi^{\dagger}i\Gamma_{200}\Psi].
\end{equation}

Lastly, the dynamics of the OP is described by the standard Ginzburg-Landau action
\begin{equation}
    S_{B}=\int d\tau d^{D}\mathbf{r}\left[\phi(-\partial^{2}_{\tau}-\nabla^2 +m_B^2)\phi +\frac{\lambda}{2}\phi^4\right],
\end{equation}
where \(m_B^2\) is the tuning parameter for the continuous phase
transition with \(m_B^2>0\) (\(m_B^2<0\)) in the $U(1)$ symmetric (symmetry broken) phase. We furthermore set the bosonic and fermionic velocities to be equal to unity in the critical region~\cite{Roy2016}, see also Sec.~\ref{App:Zv}. 

The fermionic and the bosonic propagators read, respectively,
\begin{equation}\label{fermion-prop}
\begin{aligned}[b]
    G_{F}(i\omega,\mathbf{k})=&\big{\{}i\omega A_- +B(k_x\Gamma_{001}+k_y\Gamma_{032})
    +A_-(\alpha k_x \Gamma_{330}+\Gamma_{331}k_x+\Gamma_{302}k_y)\\
    &+2\alpha k_x[\Gamma_{330}k^2+\alpha k_x(k_x\Gamma_{331}+k_y\Gamma_{302})]\big{\}}\frac{1}{A_+^2 -B^2},
\end{aligned}
\end{equation}
and 
\begin{equation}\label{boson-prop}
    G_B (i\omega, \mathbf{k})=\frac{1}{\omega^2 +k^2 +m_B^2}.
\end{equation}
where \(A_{\pm}\equiv A_\pm(i\omega,{\bf k})=-\omega^2 \pm\alpha^2 k_x^2 -k^2\) and \(B\equiv B(i\omega,{\bf k})=2\alpha i\omega k_x\).

We point out that the form of the propagator corresponding to the asymmetrically tilted system, described by the Hamiltonian in Eq~\eqref{eq:hamiltonian-symmbreaking}, preserves the form in Eq.~\eqref{fermion-prop}, since the asymmetric tilt commutes with the untilted piece of the Dirac Hamiltonian. Explicitly, the propagator of the asymmetric TDFs reads
\begin{equation}\label{fermion-prop-asymm}
\begin{aligned}[b]
    G_{F,AS}(i\omega,\mathbf{k})=&\big{\{}i\omega A_- +B(k_x\Gamma_{031}+k_y\Gamma_{002})
    +A_-(\alpha k_x \Gamma_{300}+\Gamma_{331}k_x+\Gamma_{302}k_y)\\
    &+2\alpha k_x[\Gamma_{300}k^2+\alpha k_x(k_x\Gamma_{331}+k_y\Gamma_{302})]\big{\}}\frac{1}{A_+^2 -B^2},
\end{aligned}
\end{equation}
and such a structure of the propagator, as we show in the following, implies that all the conclusions derived for the symmetric tilt equally apply to the asymmetric case. We therefore consider only the fully symmetric TDFs.

\section{Renormalization group analysis}~\label{sec:RG analysis}

Next, we proceed with the analysis of the U(1) GNY field theory. The engineering (bare) dimensions of the couplings are \(\text{dim}[g^2]=\text{dim}[\lambda]=3-D\) and for the tilt parameter \(\text{dim}[\alpha]=0\), implying that \(D=3\) is the upper critical spatial dimension of the theory. We therefore employ an \(\epsilon\)-expansion about the upper critical dimension,  \(\epsilon=3-D\), to access the quantum-critical behavior in \(D=2\). To this end, we define the Euclidean  action \(S_{R}=\int d\tau d\mathbf{r} L_{R}\), where the renormalized Lagrangian is

\begin{equation}\label{eq:renormalized-action}
\begin{aligned}[b]
L_{R}=&\Psi^{\dagger}\left[Z_{\Psi}\left(\partial_{\tau}-i Z_v v\Gamma_{331}\partial_{x} -i Z_v v\Gamma_{302}\partial_{y} -i Z_v Z_{\alpha}\alpha v\Gamma_{330}\partial_{x}\right)\right]\Psi +Z_{g}L_{Y}\\
    &+Z_{\phi}\phi(-\partial^{2}_{\tau}-\nabla^2 )\phi +Z_{m_B}m_B^2 \phi^2+Z_{\lambda}\frac{\lambda}{2}\phi^4,
\end{aligned}
\end{equation}
where \(Z_{j}\)'s are renormalization constants, and we restore the Fermi velocity for completeness. 

To make a connection with relativistic Dirac  field theories more explicit, we here rewrite the above Lagrangian in the Minkowski spacetime  by performing a Wick rotation from the imaginary to real time ($\tau\to -i t$),  set $Z_v=1$, anticipating this result (see App.~\ref{App:Zv}), and fix both fermionic and bosonic velocities to unity, which then takes a form 
\begin{equation}\label{eq:renormalized-action-relativistic}
\begin{aligned}[b]
\tilde{L}_{R}=&Z_\Psi i\bar{\Psi}\Gamma^\mu\partial_\mu\Psi -Z_{g}g\left[(\text{Re}\,\phi)\bar{\Psi}\Gamma^5\Psi-(\text{Im}\,\phi)\bar{\Psi}\Psi\right]
    +Z_{\phi}\,(\partial^\mu\phi^*)(\partial_\mu\phi)\\
    &+Z_{m_B}m_B^2 |\phi|^2+Z_{\lambda}\frac{\lambda}{2}|\phi|^4+\tilde{L}_{\rm tilt}.
\end{aligned}
\end{equation} 
Here, we defined  $\bar{\Psi}=\Psi^\dagger\Gamma^0$, with $\Gamma^0=\Gamma_{200}$, while $\Gamma^1=-i\Gamma_{131}$, $\Gamma^2=-i\Gamma_{102}$, $\Gamma^5=\Gamma_{300}$, and the matrix $\Gamma^3=-i\Gamma_{133}$ can be used to extend the Dirac kinetic term to the $(3+1)-$dimensional Minkowski space. 
This is so because  these five $\Gamma$ matrices,  acting on the (complex) Dirac fermions,  form a Clifford algebra of mutually anticommuting matrices. Furthermore, the matrix $\Gamma_5$ can be taken as the generator of chiral symmetry, which is  inherited  from the particle-hole symmetry in the Euclidean theory, while the valley degrees of freedom become trivial in this representation (all the five $\Gamma-$matrices are diagonal in the valley index). 
Finally, the Lagrangian of the tilt term, $\tilde{L}_{\rm tilt}$, which can be considered as a minimal Lorentz-symmetry breaking deformation of the above fully relativistic Dirac field theory, reads 
\begin{equation}
\tilde{L}_{\rm tilt}= i Z_\alpha\alpha \bar{\Psi}\Gamma_{\rm tilt}\partial_x \Psi, 
\end{equation}
with the matrix $\Gamma_{\rm tilt}=-i\Gamma_{130}$.

\subsection{$\beta-$functions for a tilted Dirac $s-$wave superconductor}

To obtain RG flow of the couplings, we integrate out the fermionic and bosonic  modes with frequency \(-\infty <\omega<\infty\) and use dimensional regularization in \(D=3-\epsilon\) spatial dimensions within the minimal subtraction scheme to handle the ensuing divergences. The one-loop computation of the self-energy diagrams for the fermions (App.~\ref{app:fermion-self-energy}) and for the bosonic order parameter (App.~\ref{app:bosonic-SE}) then yields the fermion field renormalization (App.~\ref{app:Zpsi}), the boson field renormalization (\ref{app:bosonic-SE}), and the renormalization constant for the tilt parameter (App.~\ref{app:Zalpha}), which, respectively, read
\begin{equation}\label{ZPsi-main}
    Z_{\Psi}=1-\frac{g^2}{(4\pi)^2}G(\alpha)\frac{1}{\epsilon},
\end{equation}

\begin{equation}\label{eq:Zphi-main}
Z_{\Phi}=1-\frac{g^2}{4 \pi^2}N_F F(\alpha)\frac{1}{\epsilon}
\end{equation}

\begin{equation}\label{eq:Zalpha-main}
    Z_{\alpha}=1+\frac{g^2}{(4\pi)^2}\left[G(\alpha)-I(\alpha)\right]\frac{1}{\epsilon},
\end{equation}
while the renormalization constant for the velocity $Z_v=1$, as shown in App.~\ref{App:Zv}. Here, the  functions are defined as 
$F(\alpha)=(1-\alpha^2)^{-2}$, $G(\alpha)=(1-(\alpha/2)^2)^{-1}$, while  $I(\alpha)$ is given by Eq.~\eqref{eq:I}. 

The computation of the vertex diagrams to the same order gives the renormalization conditions for the Yukawa coupling constant (App.~\ref{App:Yukawa vertex}) 
\begin{equation}\label{eq:g2-condition}
Z_{\phi}^{1/2}\,Z_{\Psi}\,g_0\, \mu^{-\epsilon/2}=g,
\end{equation}
and for for the $\Phi^4$ coupling (App.~\ref{App4})
\begin{equation}\label{eq:cond-lambda}
    Z_{\phi}^2 \lambda_0 \mu^{-\epsilon}-\left((N_B +4 )\frac{\lambda^2}{8\pi^2}-\frac{g^4}{2\pi^2}N_F F(\alpha)\right)\frac{1}{\epsilon}=\lambda.
\end{equation}
Here, the couplings with (without) the subscript are the bare (renormalized) ones. Furthermore, \(\mu\) is the RG (momentum) scale, and we consider  \(N_B\) complex components for the bosonic OP.

\begin{figure}[t!]
     \centering
     \begin{subfigure}[b]{0.32\textwidth}
         \centering
         \includegraphics[width=\textwidth]{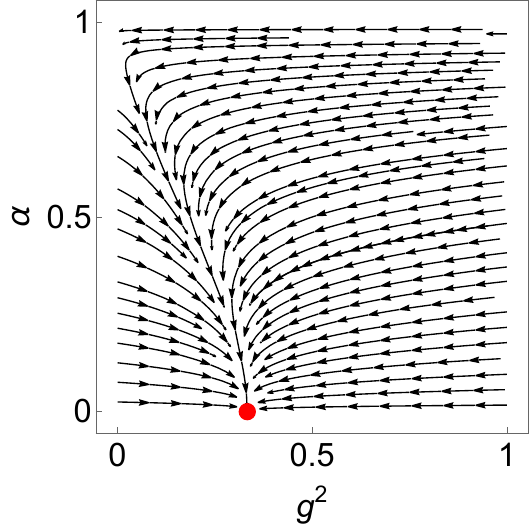}
         \caption{}
         \label{fig:y equals x}
     \end{subfigure}
     \begin{subfigure}[b]{0.32\textwidth}
         \centering
         \includegraphics[width=\textwidth]{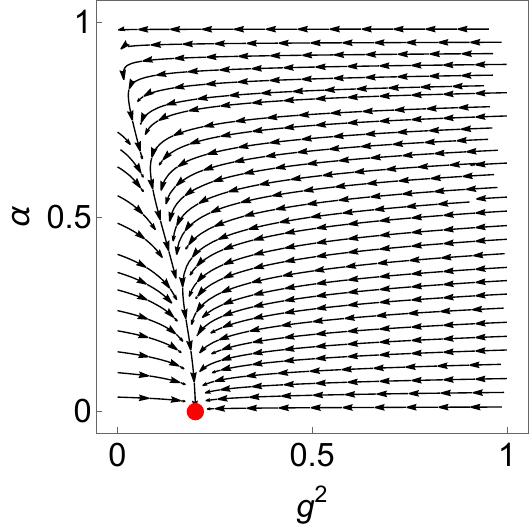}
         \caption{}
         \label{fig:three sin x}
     \end{subfigure}
     \begin{subfigure}[b]{0.32\textwidth}
         \centering
         \includegraphics[width=\textwidth]{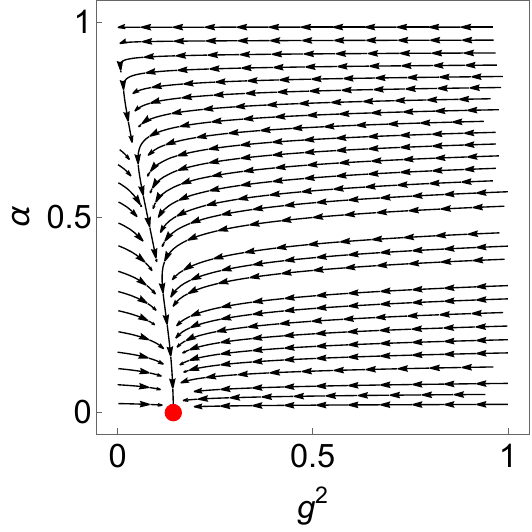}
         \caption{}
         \label{fig:five over x}
     \end{subfigure}
        \caption{One-loop RG flow  of the tilt parameter \(\alpha\) and the Yukawa coupling \(g^2\) at fixed \(\epsilon=1\) in the case of the $s-$wave superconducting instability. The red dot corresponds to the Gross-Neveu-Yukawa superconducting quantum-critical point (QCP), for different number of eight-component Dirac flavors ($N_F$). RG flow is shown for  (a)  \(N_F =1\);  (b)  \(N_F =2\); (c)  \(N_F =3\). The analytical form of the Lorentz-symmetric QCP ($\alpha=0$) is given by Eq.~\eqref{eq:SC-QCP}.}
        \label{fig:RGflow}
\end{figure}

Using  the above renormalization conditions, we obtain the RG \(\beta\) functions to the leading order in the \(\epsilon\)-expansion, in the critical hyperplane \(m_B^2=0\), see Eq.~\eqref{eq:renormalized-action}. We define the $\beta$ function for a coupling constant $X$ as \(\beta_X\equiv -dX/d\ln{(\mu)}|_{\{X_0\}}\), with all the bare couplings ($\{X_0\}$) fixed. Since the renormalization constant for the velocity $Z_v=1$, the renormalization condition $Z_v v=v_0$ implies that the Fermi velocity is marginal, $\beta_v=0$. Furthermore, the renormalization condition for the tilt term, $Z_\alpha\alpha=\alpha_0$ implies that
\begin{equation}\label{eq:beta-alpha}
    \beta_\alpha=-\alpha \frac{g^2}{2}[G(\alpha)-I(\alpha)]\equiv -\alpha g^2 K(\alpha),
\end{equation}
\begin{figure}[t]
    \centering
\includegraphics[width=0.6\textwidth]{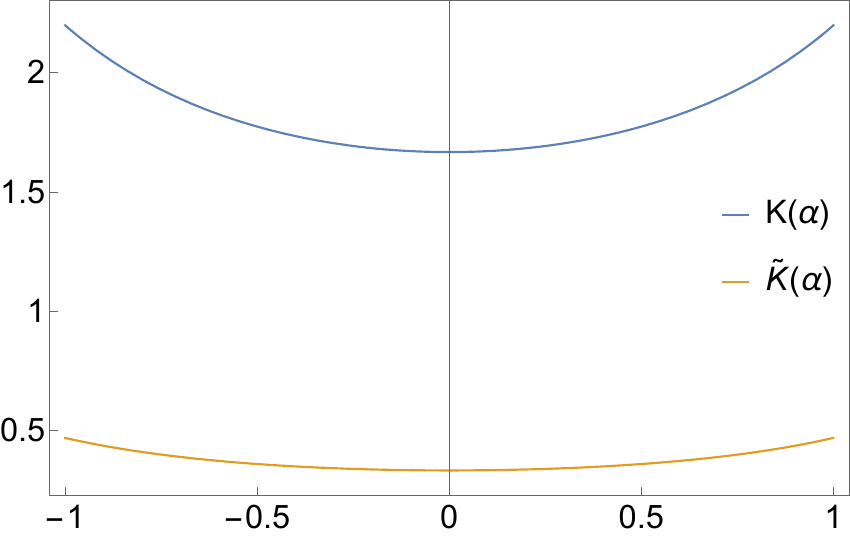}
    \caption{The form of the functions \(K(\alpha)\) and  
    \({\tilde K}(\alpha)\) governing the renormalization-group flow of the tilt parameter at the transition from the tilted Dirac semimetal to the $s-$wave superconducting (SC) and the charge-density wave (CDW) phases, respectively. The corresponding  $\beta$ function is given by Eq.~\eqref{eq:beta-alpha} [Eq.~\eqref{eq:beta-alpha-CDW}] for the $s-$wave SC [CDW] instability.} 
    \label{Fig:Kalpha}
\end{figure}
with the function $K(\alpha )$ as  strictly positive for $|\alpha|<1$, as shown in Fig.~\ref{Fig:Kalpha}. Therefore, the tilt parameter is \emph{irrelevant}, and  flows to zero as the system approaches the critical point governing the transition into the $s$-wave superconducting state. Ultimately, the Lorentz symmetry is restored at the QCP  and the quantum-criticality is controlled by the GNY QCP for $\alpha=0$. To confirm this scenario, using the renormalization condition in Eq.~\eqref{eq:g2-condition}, we furthermore find the $\beta-$function for the Yukawa coupling
\begin{equation}\label{eq:beta-g2}
    \beta_{g^2}=\epsilon g^2 - [2N_F F(\alpha) + G(\alpha)]g^4 ,
\end{equation}
where we rescaled  \(g^2 /8\pi^2 \rightarrow g^2\).
The RG flow in the $(g^2,\alpha)$ plane is shown in Fig.~\ref{fig:RGflow}, which indeed corroborates the irrelevance of the tilt parameter, and the ensuing restoration of the Lorentz symmetry at the nontrivial QCP with the value given by Eq.~\eqref{eq:beta-g2}, 
\begin{equation}
g^2_\star=\frac{\epsilon}{[2N_F F(0) + G(0)]}.
\end{equation}

For completeness, we also show the RG flow equation for the $\Phi^4-$coupling for the tilted Dirac superconductor, 
\begin{equation}\label{eq:beta-lambda}
\beta_{\lambda}=\epsilon\lambda -(4+N_B)\lambda^2 +4N_F g^2 F(\alpha)(g^2-\lambda),
\end{equation}
which is obtained from the renormalization condition in Eq.~\eqref{eq:cond-lambda}, after rescaling $X/8\pi^2\rightarrow X$, with $X=\{g^2,\lambda\}$. Notice that  the $\beta$ functions in Eqs.~\eqref{eq:beta-g2} and~\eqref{eq:beta-alpha} for the untilted Dirac fermion ($\alpha=0$) coincide with the ones obtained in Ref.~\cite{RJH-PRB2013}. The corresponding fixed points of the RG flow are obtained  from the $\beta-$functions for the Yukawa and the $\lambda$ couplings after taking $\alpha=0$, as implied by the irrelevance of the tilt term [Eq.~\eqref{eq:beta-alpha}]. Only one of them is stable in the critical plane ($m_B^2=0$), and therefore represents the critical point, which is located at~\cite{RJH-PRB2013} 
\begin{equation}~\label{eq:SC-QCP}
(g_\star^2,\lambda_\star)=\frac{\epsilon}{X+4N_F}\left(1,X+\sqrt{X^2+8(1-X)(4+N_B)}\right),
\end{equation}
where $X=1-2N_F$. Notice that $\lambda_\star$ is positive, and therefore corresponds to a second-order phase transition,  only for $N_F>1/2$, or equivalently, in the Dirac system featuring at least one eight-component Dirac fermion flavor. 

We emphasize here that identical renormalization factors and vertex corrections are obtained for the asymmetric TDFs,  since the tilt matrix in this case anticommutes with the SC OP and commutes with the untilted part of the Dirac Hamiltonian, which is the same as for the symmetric TDFs, while all the form factors in the propagators for the asymmetric [Eq.~\eqref{fermion-prop-asymm}] and symmetric [Eq.~\eqref{fermion-prop}] TDFs are identical. 


\subsection{Comparison with charge-density-wave instability}
\begin{figure}[t!]
     \centering
     \begin{subfigure}[b]{0.32\textwidth}
         \centering
         \includegraphics[width=\textwidth]{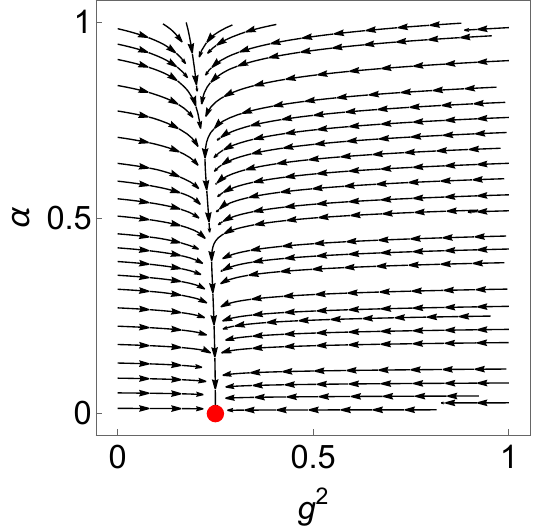}
         \caption{}
         \label{fig:y equals x}
     \end{subfigure}
     \begin{subfigure}[b]{0.32\textwidth}
         \centering
         \includegraphics[width=\textwidth]{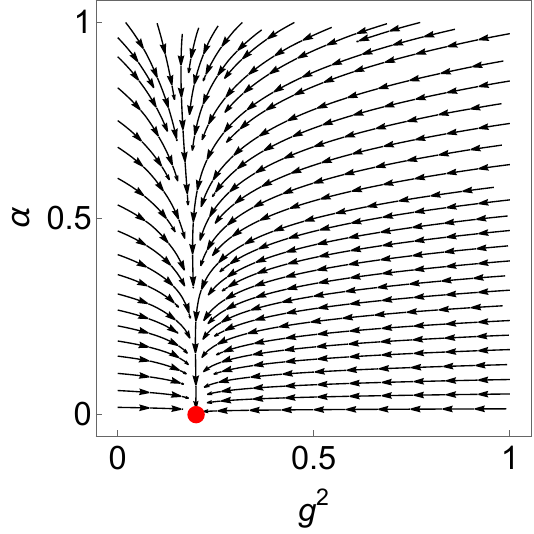}
         \caption{}
         \label{fig:three sin x}
     \end{subfigure}
     \begin{subfigure}[b]{0.32\textwidth}
         \centering
         \includegraphics[width=\textwidth]{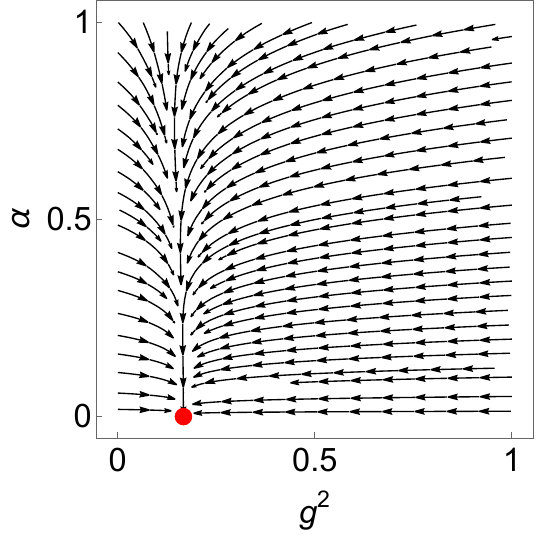}
         \caption{}
         \label{fig:five over x}
     \end{subfigure}
        \caption{One-loop RG flow  of the tilt parameter \(\alpha\) and the Yukawa coupling \(g^2\) at fixed \(\epsilon=1\) in the case of the charge-density-wave (CDW) instability. The red dot corresponds to the Gross-Neveu-Yukawa CDW quantum-critical point (QCP), for different number of eight-component Dirac flavors ($N_F$). RG flow is shown for  (a)  \(N_F =1\);  (b)  \(N_F =2\); (c)  \(N_F =3\). The analytical form of the Lorentz-symmetric QCP ($\alpha=0$) is given by Eq.~\eqref{eq:CDW-QCP}.}
        \label{fig:RGflowCDW}
\end{figure}

We here use the opportunity to revise the problem of a charge-density wave (CDW) instability breaking the sublattice symmetry of a tilted  Dirac semimetal~\cite{Rostami-PRR2020}, in light of the restoration of the Lorentz symmetry at the superconduting QCP. The CDW OP takes the form, 
\begin{equation}
\Phi(\mathbf{r},\tau)=\left\langle{\Psi^{\dagger}(\mathbf{r},\tau)\Gamma_{303}\Psi(\mathbf{r},\tau)}\right\rangle,
\end{equation}
implying that the Yukawa action reads  
\begin{equation}
 S_{Y}=g\int d\tau d^{D}\mathbf{r}
\,\,\Phi\,\,\Psi^{\dagger}\Gamma_{303}\Psi, 
\end{equation}
with the OP matrix  anticommuting  with the untilted part of the Dirac Hamiltonian in Eq.~\eqref{hamiltonian}, and therefore gaps it out. 
We are interested here only in the RG flow of the tilt parameter and the Yukawa coupling, and therefore consider the fermionic self-energy, which reads 
\begin{equation}\label{Fselfenergy-CDW}
\begin{aligned}[b]
\Sigma_{F}(i\Omega,{\bf k})=&g^2\int[d\omega]\int[d^{D}\mathbf{q}]\,\Gamma_{303}\, G_{F}(i\omega,\mathbf{q})\,\Gamma_{303}\, G_{B}(i\Omega-i\omega,\mathbf{q})\\
    =& g^2\int[d\omega]\int[d^{D}\mathbf{q}]\,\left[i\omega A_- -B(q_x\Gamma_{001}+q_y\Gamma_{032})
    +\alpha q_x (A_-+2q^2)\Gamma_{330} \right.\\
    &\left.-(A_- +2\alpha^2q_x^2)(q_x\Gamma_{331}+q_y\Gamma_{302})\right]\frac{1}{[A_+^2+B^2][(\omega-\Omega)^2-({\bf q}-{\bf k})^2]},
\end{aligned}
\end{equation}
which is obtained directly from the self-energy for the SC OP, Eq.~\eqref{app:fermion-self-energy}, taking into account that the CDW OP commutes with the tilt matrix, $\Gamma_{330}$. The functions $A_\pm$ and $B$ are defined in Eq.~\eqref{hamiltonian}.  This form of the fermionic self-energy, in turn, yields the renormalization constant for the field ($Z_\Psi$) in the form given by Eq.~\eqref{ZPsi-main}, while the factor $Z_{\alpha}$ reads 
\begin{equation}\label{eq:Zalpha-main-CDW}
    Z_{\alpha}=1+\frac{g^2}{(4\pi)^2}\left[G(\alpha)+I(\alpha)\right]\frac{1}{\epsilon},
\end{equation}
with the function $G(\alpha)$ [$I(\alpha)$] defined in Eq.~\eqref{eq:Zalpha-main}  [Eq.~\eqref{eq:I}]. The corresponding RG flow equation  for the tilt is then straightforwardly obtained from the renormalization condition $Z_\alpha\alpha=\alpha_0$, yielding 
\begin{equation}\label{eq:beta-alpha-CDW}
    \beta_\alpha=-\alpha \frac{g^2}{2}[G(\alpha)+I(\alpha)]\equiv -\alpha g^2 {\tilde K}(\alpha),
\end{equation}
with the strictly positive function ${\tilde K}(\alpha)$ shown in Fig.~\ref{Fig:Kalpha}. Therefore, the Lorentz symmetry gets restored for the CDW instability, but much slower than in the SC case, since ${\tilde K}(\alpha)<K(\alpha)$, for $|\alpha|<1$, as can be seen in Fig.~\ref{Fig:Kalpha}. 
Furthermore, the flow of the Yukawa coupling in this case is given by 
\begin{equation}~\label{eq:g2-cdw}
\beta_{g^2}=\epsilon g^2-(N_F+3)G(\alpha)g^4,
\end{equation}
with the  stable QCP realized at 
\begin{equation}~\label{eq:CDW-QCP}
    g_\star^2=\frac{\epsilon}{(3+N_F)G(0)}. 
\end{equation}
The corresponding RG flow in the $(g^2,\alpha)$ plane is displayed in Fig.~\ref{fig:RGflowCDW}, which corroborates the restoration of the Lorentz symmetry at the QCP governing the transition into the CDW phase, and making the line of critical points found in Ref.~\cite{Rostami-PRR2020} eventually unstable. 

We point out that the sign in front of the strictly negative function $I(\alpha)$ (for $|\alpha|<1$) in Eq.~\eqref{eq:beta-alpha-CDW},  is a direct consequence of the \emph{commutation} of the tilt matrix ($\Gamma_{330}$)  and the CDW OP $\Gamma_{303}$, which is diagonal in the Nambu space. This overall \emph{positive} contribution   is responsible for slowing down the RG flow of the tilt with respect to the SC case, implying that the restoration of the Lorentz symmetry takes longer RG time. 

In general, we can observe that  the rate of the RG flow for the tilt parameter, and therefore the  rate of the Lorentz symmetry restoration is directly related to the (anti)commutation relation  of the OP with the tilt matrix. Namely,  since all the $\Gamma-$matrices square to unity and either commute or anticommute with each other,  an OP that commutes with the tilt matrix is expected to make the tilt parameter \emph{less irrelevant} than an anticommuting OP. In particular, an insulating (particle-hole) instability diagonal in other flavor subspaces (valley, sublattice, etc.) should render the tilt less irrelevant than a superconducting (particle-particle) instability, as the examples of the CDW and the $s-$wave instabilities explicitly demonstrate. Finally, the form of the propagator of the asymmetric TDFs implies that the $\beta-$functions at CDW QCP for the asymmetric TDFs are of the same form  as for the symmetric TDFs.

\section{Conclusions and outlook}~\label{sec:conclusions}
In this paper, we have shown that the Lorentz symmetry is restored at the strongly-coupled QCP separating the tilt Dirac semimetal from a gapped $s-$wave SC instability through the irrelevance of the tilt operator and its ultimate vanishing at the QCP. As we further argued such a symmetry restoration is a quite generic feature for the strongly interacting TDFs, irrespective of whether they break a discrete mirror lattice symmetry at the noninteracting level, i.e. whether they feature symmetric [Eq.~\eqref{hamiltonian}] or asymmetric [Eq.~\eqref{eq:hamiltonian-symmbreaking}] tilt.  

Our results can be routinely tested in the lattice quantum Monte Carlo simulations~\cite{Sorella1992} of the Hubbard interacting TDFs. In light of this, we emphasize that even though the tilt parameter is irrelevant, and ultimately vanishes at the QCP, it may take  long RG time to reach this value.  Therefore in any finite-size system the effects of a finite tilt may still be appreciable, as given by the solution of the flow equations~\eqref{eq:beta-alpha},~\eqref{eq:beta-g2} and~\eqref{eq:beta-lambda}. See also Fig.~\ref{fig:RGflow}. 

Directly pertinent to the SC instability of the TDFs is whether the coupling to a dynamical $U(1)$ gauge field can lead to the breaking of the Lorentz symmetry, which is an issue that we plan to investigate in future. In particular, it would be interesting to see whether such a coupling could help stabilization of a charged QCP at a finite coupling. 

Our work should motivate further studies of the interaction effects in the non-Hermitian tilted Dirac materials with the coupling to the environment explicitly taken into account through non-Hermitian terms in the Hamiltonian-like operator. In particular, it will be important to address whether the  restoration of the Lorentz symmetry is also featured in the non-Hermitian tilted Dirac semimetals, similarly to the case of  untilted non-Hermitian Dirac fermions~\cite{juricic2023yukawalorentz}.

Finally, we would like to point out that these results may be relevant for future studies of the strong coupling instabilities of type-2 Dirac semimetals with  electron- and hole Fermi surfaces, and should spur further research efforts in this direction.

\acknowledgments

We thank Bitan Roy for fruitful discussions. This work was supported by  the Swedish Research Council Grant No. VR 2019-04735 (V.J.),  Fondecyt (Chile) Grant No. 1230933 (V.J.) and ANID/ACT210100 (V.J.).

\appendix

\section{Renormalization group analysis the Gross-Neveu-Yukawa theory}\label{A}

In this appendix we present the details of the renormalization group calculation to the one-loop order within the $\epsilon=3-D$ expansion about $D=3$ upper critical dimension of the quantum-critical theory describing interacting  TDFs. The short-range interaction is encoded through the Yukawa coupling between the Dirac fermion quasiparticles and the bosonic degrees of freedom representing the fluctuations of the $s$-wave superconducting  phase 
\begin{equation}
    S_{Y}=g\int d\tau d^{D}\mathbf{r}[(\text{Re}\phi)\Psi^{\dagger}\Gamma_{100}\Psi+(\text{Im}\phi)\Psi^{\dagger}i\Gamma_{200}\Psi]\equiv \int d\tau d^{D}\mathbf{r}\, L_{Y}.
\end{equation}
It is convenient to rewrite the coupling of the bosonic order parameter to the fermionic field explicitly in terms of the real and imaginary parts of the OP,
\begin{equation}
L_{Y}=g\left[(\text{Re}\phi)\Psi^{\dagger}\Gamma_{100}\Psi+(\text{Im}\phi)\Psi^{\dagger}i\Gamma_{200}\Psi\right]= g\left[\phi \Psi^{\dagger}P_{+}\Psi+\phi^{*}\Psi^{\dagger}P_{-}\Psi\right],
\end{equation}
where the matrices $P_\pm$ are defined as 
\begin{equation}\label{eq:projectors}
P_{\pm}=\frac{1}{2}\left(\Gamma_{100}\mp i\Gamma_{200}\right),
\end{equation}
and we employ  eight-dimensional representation of the $\Gamma-$matrices.

To address the quantum-critical behavior in $D=2$,  we employ \(\epsilon=3-D\),  the deviation from the upper critical $D=3$ spatial dimensions,  as an expansion parameter.

\begin{figure}[t!]
     \centering
     \begin{subfigure}[b]{0.25\textwidth}
         \centering
         \includegraphics[width=\textwidth]{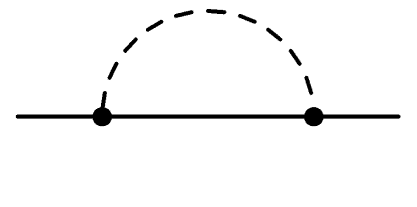}
         \caption{}
         \label{fig:Fselfenergy}
     \end{subfigure}
     \begin{subfigure}[b]{0.25\textwidth}
         \centering
         \includegraphics[width=\textwidth]{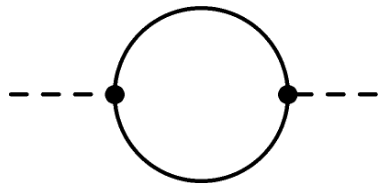}
         \caption{}
         \label{fig:Diagramb}
     \end{subfigure}
     \begin{subfigure}[b]{0.25\textwidth}
         \centering
         \includegraphics[width=\textwidth]{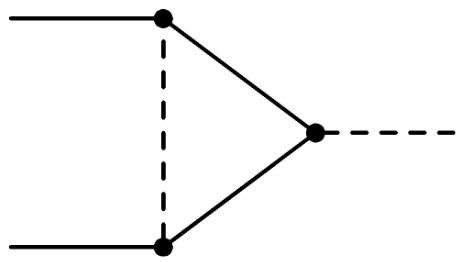}
         \caption{}
         \label{fig:diagramc}
         
     \end{subfigure}
     \begin{subfigure}[b]{0.25\textwidth}
         \centering
         \includegraphics[width=\textwidth]{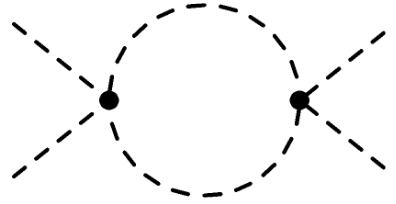}
         \caption{}
         \label{fig:diagramd}
         
     \end{subfigure}
     \begin{subfigure}[b]{0.25\textwidth}
         \centering
         \includegraphics[width=\textwidth]{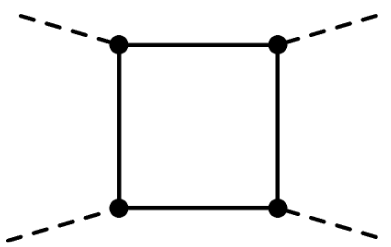}
         \caption{}
         \label{fig:diagrame}
         
     \end{subfigure}
        \caption{Feynman diagrams to the leading order in the \(\epsilon\) expansion: (a) fermionic self-energy, (b) bosonic self-energy, (c) correction to the Yukawa vertex arising from fermion excitations, (d) correction to the \(\lambda\) vertex from the boson self-interaction, and (e) correction to the \(\lambda\) vertex from the Yukawa interaction. The solid line represents the fermionic propagator, while the dashed line stands for the bosonic propagator.}
        \label{fig:Diagrams}
\end{figure}

\subsection{Fermionic self energy}\label{app:fermion-self-energy}

We compute to the leading-order the zero-temperature (\(T=0\)) self-energy for tilted Dirac fermions (Fig.~ \ref{fig:Fselfenergy}) in the critical hyperplane (all masses set to zero)
\begin{equation}
    \Sigma_{F}(i\Omega,\mathbf{k})=g^2\int[d\omega]\int[d^{D}\mathbf{q}]\{P_{+}G_{F}(i\omega,\mathbf{q})P_{-}+P_{-}G_{F}(i\omega,\mathbf{q})P_{+}\}G_{b}(i\Omega-i\omega,\mathbf{q}-\mathbf{k})
\end{equation}
where \([d\omega]=d\omega/2\pi\) and \([d^{D}\mathbf{q}]=d^D \mathbf{q}/(2\pi)^D\). The bosonic and fermionic propagators are given in the main text, Eqs.~\eqref{fermion-prop} and \eqref{boson-prop}. The explicit form of the fermionic self-energy is
\begin{equation}\label{Fselfenergy}
\begin{aligned}[b]
    \Sigma_{F}(i\Omega,{\bf k})=&g^2\int[d\omega]\int[d^{D}\mathbf{q}]\{P_{+}G_{F}(i\omega,\mathbf{q})P_{-}+P_{-}G_{F}(i\omega,\mathbf{q})P_{+}\}G_{B}(i\Omega-i\omega,\mathbf{q})\\
    =& g^2\int[d\omega]\int[d^{D}\mathbf{q}]\{P_{+},P_{-}\}\left[i\omega A_- +B(q_x\Gamma_{001}+q_y\Gamma_{032})-\alpha q_x (A_-+2q^2)\Gamma_{330} \right.\\
    &\left.-(A_- +2\alpha^2q_x^2)(q_x\Gamma_{331}+q_y\Gamma_{302})\right]\frac{1}{(\omega^2-\alpha^2 q_x^2+q^2)^2+(2\alpha \omega q_x)^2}\\
    &\times \frac{1}{(\omega-\Omega)^2-({\bf q}-{\bf k})^2},
\end{aligned}
\end{equation}
with the anticommutator $\{P_+,P_-\}=1$, as can be directly obtained from  Eq.~\eqref{eq:projectors};  
\(A_{\pm}\equiv A_\pm(i\omega,{\bf k})=-\omega^2 \pm\alpha^2 k_x^2 -k^2\) and \(B\equiv B(i\omega,{\bf k})=2\alpha i\omega k_x\). We here also set Fermi velocity $v=1$, since it does not get renormalized due to the relativistic invariance of the Yukawa interaction [Eq.~\eqref{eq:Yukawa}], as also explicitly shown in App.~\ref{App:Zv}.  
The algebraic structure of the integrand  allows us to systematically calculate the corresponding renormalization constants ($Z's$), which we explicitly show in the following. Notice that the negative signs multiplying the last two terms in the above self-energy result from the anticommutation of the OP matrices ($P_\pm$) with the tilt matrix $\Gamma_{330}$. 

\subsubsection{Fermionic field renormalization ($Z_\Psi$)}\label{app:Zpsi}

For the renormalization constant \(Z_{\Psi}\), only the terms proportional to the unity matrix contribute.  Thus after setting the external momentum to zero,  ${\bf k}=0$, and   performing the integral over \(\omega\), we obtain 
\begin{equation}
\Sigma_{F,000}(i\Omega,0)=-g^2\int[d^{D}\mathbf{q}]\frac{i \Omega  \left(4 q^2+\alpha ^2 q_{x}^2+\Omega ^2\right)}{2 q \left(16 q^4+8 q^2 \left(\Omega ^2-\alpha ^2 q_{x}^2\right)+\left(\alpha ^2 q_{x}^2+\Omega ^2\right)^2\right)}.
\end{equation}
Now, to find the fermion field renormalization $Z_\Psi$, we take
\begin{equation}
    \lim_{i\Omega\to 0}\frac{\partial\Sigma_{F,000}(i\Omega,0)}{\partial (i\Omega)}=-\frac{g^2}{(4\pi)^2}\,G(\alpha)\int_{\lambda}^{\Lambda} dq\frac{1}{q},
\end{equation}
with \(\lambda=\Lambda/b\), where the Wilsonian RG parameter  \(b>1\), and \(b-1\ll1\). Then, we use the correspondence between the hard cutoff and dimensional regularization
\begin{equation}
\int_{\lambda}^{\Lambda}\frac{dq}{q}\rightarrow \frac{1}{\epsilon},
\end{equation}
and the function \(G(\alpha)\) reads
\begin{equation}\label{Galpha}
G(\alpha)=\int_{0}^{2\pi}\frac{d\phi}{\pi}\int_{0}^{\pi}d\theta \sin(\theta)\frac{\left(4+\alpha ^2 \sin ^2(\theta ) \cos ^2(\phi )\right)}{ \left(4- \alpha ^2 \sin ^2(\theta ) \cos ^2(\phi )\right)^2}\\
=\frac{4}{4-\alpha^2},
\end{equation}
Notice that the function $G(\alpha)$ is strictly positive for \(|\alpha|<1\). We then obtain the renormalization constant for the fermion field through the following renormalization condition
\begin{equation}\label{Zpsi}
    Z_{\Psi}=1+\lim_{i\Omega\to 0}\frac{\partial\Sigma_{f,000}(i\Omega,0)}{\partial i\Omega}=1-\frac{g^2}{(4\pi)^2}G(\alpha)\frac{1}{\epsilon}.
\end{equation}


\subsubsection{Renormalization factor for the tilt parameter ($Z_\alpha$)}\label{app:Zalpha}

The renormalization constant for the tilt parameter, $Z_\alpha$, is obtained from the term in the fermionic self-energy proportional to the matrix $\Gamma_{330}$, as dictated by the form of the tilt term in the bare theory [see Eq.~\eqref{hamiltonian}]. This, together with the form of the self-energy, in turn, dictates the following renormalization condition for the tilt, 
\begin{equation}\label{AlphaCondition}
    \left.-Z_{\Psi}Z_{\alpha}\alpha k_x  - k_x \frac{\partial{\Sigma}_{F,330} (i\Omega,\mathbf{k})}{\partial k_x}\right|_{i\Omega \to 0, \mathbf{k}\to 0}=-\alpha k_x ,
\end{equation}
where ${\Sigma}_{F,330} (i\Omega,\mathbf{k})$ is the term proportional to the matrix $\Gamma_{330}$ in the fermionic self-energy in Eq.~\eqref{Fselfenergy},
\begin{equation}\label{App:Sigmatilde}
    {\Sigma}_{F,330} (i\Omega,\mathbf{k})=g^2\int[d\omega]\int[d^{D}\mathbf{q}]\frac{\alpha q_x(\omega^2+\alpha^2q_x^2-q^2)}{[(\omega^2 - \alpha^2 q_x^2 +q^2)^2+(2\alpha \omega q_x)^2][(\omega-\Omega)^2-({\bf q}-{\bf k})^2]}.
\end{equation}
We notice that we here explicitly used marginality of the Fermi velocity $v$ in Eq.~\eqref{hamiltonian}, due to the relativistic invariance of the Yukawa term, see Eq.~\eqref{eq:Yukawa}.

We then take a finite external momentum $\mathbf{k}=k_x {\bf e}_x$ in Eq.~\eqref{App:Sigmatilde}, and keep  only the terms proportional to \(\Gamma_{330}\), which, by virtue of the renormalization condition in Eq.~\eqref{AlphaCondition}, yields,
\begin{equation}
   Z_\alpha Z_\Psi=1-\frac{g^2}{(4\pi)^2\epsilon}I(\alpha),
\end{equation}
where  
\begin{equation}\label{eq:I}
    I(\alpha)=\int_0^{2\pi}\frac{d\phi}{\pi}\int_0^\pi d\theta\,\,\frac{\cos^2\phi\sin^3\theta[-8+\alpha^2\cos^2\phi\sin^2\theta]}{(4-\alpha^2\cos^2\phi\sin^2\theta)^2}. 
\end{equation}
Finally, using the form of the fermion-field renormalization $Z_\Psi$ in Eq.~\eqref{Zpsi}, we find 
\begin{equation}
    Z_\alpha=1+\frac{g^2}{(4\pi)^2\epsilon}\left[G(\alpha)-I(\alpha)\right],
\end{equation}
with the function $G(\alpha)$ given by Eq.~\eqref{Galpha}, which is identical to Eq.~\eqref{eq:Zalpha-main} in the main text. 

\subsubsection{Renormalization factor for the Fermi velocity ($Z_v$)}\label{App:Zv}

We here show that the renormalization factor of the Fermi velocity, $Z_v$, is equal to unity, as a consequence of the Lorentz invariance of the Yukawa vertex in Eq.~\eqref{eq:Yukawa}. Therefore, the Fermi velocity remains marginal, as is the case in the tilt-free case. To show this feature, we consider only one component of the velocity, taken to be along the $x-$axis for concreteness. The corresponding renormalization condition reads 
\begin{equation}
\left. Z_\Psi Z_v+\frac{\partial\Sigma_{F,331}(i\Omega,\mathbf{k})}{\partial k_x}\right|_{i\Omega \to 0, \mathbf{k}\to 0}=1,
\end{equation}
where the term $\Sigma_{f,331}$ is the part of the self-energy in Eq.~\eqref{Fselfenergy} multiplied by the matrix $\Gamma_{331}$, which explicitly reads 
\begin{equation}
    \Sigma_{F,331}(i\Omega,\mathbf{k})=g^2\int [d\omega]\int [d^D\mathbf{k}]\,\,\frac{q_x(\omega^2+q^2-\alpha^2q_x^2)}{[(\omega^2+q^2-\alpha^2q_x^2)^2+(2\alpha\omega q_x)^2][(\omega-\Omega)^2+(\mathbf{q}-\mathbf{k})^2]}.
\end{equation}
An explicit calculation then yields
\begin{equation}\label{eq:Renorm-v}
    \left. \frac{\partial\Sigma_{F,331}(i\Omega,\mathbf{k})}{\partial k_x}\right|_{i\Omega \to 0, \mathbf{k}\to 0}=\frac{g^2}{2(2\pi)^3\epsilon}\int_0^{2\pi}d\phi\int_0^\pi d\theta \,\,\frac{\cos^2\phi\sin^3\theta(12-\alpha^2\cos^2\phi\sin^2\theta)}{(4-\alpha^2\cos^2\phi\sin^2\theta)^2},
\end{equation}
which, after performing the angular integrals, yields 
\begin{equation}
    \left. \frac{\partial\Sigma_{F,331}(i\Omega,\mathbf{k})}{\partial k_x}\right|_{i\Omega \to 0, \mathbf{k}\to 0}=\frac{g^2}{(4\pi)^2}G(\alpha)\frac{1}{\epsilon},
\end{equation}
with the function $G(\alpha)$ given by Eq.~\eqref{Galpha}.  
Using Eq.~\eqref{Zpsi} and the renormalization condition~\eqref{eq:Renorm-v} , we obtain that $Z_v=1$, implying that the Fermi velocity remains marginal (unrenormalized) for the TDFs to the leading order in the $\epsilon-$expansion, once the velocity isotropy is reached. Notice that we here considered only the $x-$component of the velocity, but the same conclusion holds for other components after invoking the irrelevance of the tilt parameter. 

\subsection{Bosonic self energy}\label{app:bosonic-SE}

In order to obtain the RG equations for the Yukawa coupling, we compute the self energy for the bosonic field and the correction to the Yukawa vertex. First, we compute the bosonic self energy (Fig.~\ref{fig:Diagramb}), which reads

\begin{equation}
    \Sigma_{B}(i\Omega,\mathbf{k})=-{g^2}\int[d\omega]\int[d^D \mathbf{q}]\,\text{Tr}\left[P_{+}G_{F}(i\omega+i\Omega,\mathbf{q+k})P_{-}G_{F}(i\omega,\mathbf{q})\right],
    \end{equation}
with the negative sign arising from the trace over the fermionic fields, and we set external momentum to zero (\(\mathbf{k}=0\)). After taking the trace, the integral is of the form
\begin{equation}
\begin{aligned}[b]
    \Sigma_{B}(i\Omega,0)=& -4{g^2 N_F}\int[d\omega]\int [d^{D}\mathbf{q}]\,\, \Big{(}   4k_x^4k^2 \alpha^4 - B(\omega,\mathbf{q}) B(\omega+\Omega,\mathbf{q})q^2\\
    &+4 k_x^2k^2 \alpha^2 ( A_{-}(\omega+\Omega,\mathbf{q}) +k^2) 
    +4k_x^2k^2 \alpha^2A_{-}(\omega,\mathbf{q}) \\
    &+A_{-}(\omega,\mathbf{q})A_{-}(\omega+\Omega,\mathbf{q})(k^2 +\alpha^2k_x^2+\omega(\omega+\Omega))\Big{)}\\
    &\times \Bigg(\frac{1}{[A_{+}(\omega,\mathbf{q})]^2 -B^2(\omega,\mathbf{q})}\Bigg)\Bigg(\frac{1}{[A_{+}(\omega+\Omega,\mathbf{q})]^2 -B^2(\omega+\Omega,\mathbf{q})}\Bigg),
\end{aligned}
\end{equation}
with $N_F$ as the number of eight-component Dirac fermion flavors. Taking the derivative, after performing the integral over the frequency and the momentum,  we obtain
\begin{equation}
    \lim_{\Omega\to 0}\frac{\partial\Sigma_{B}(i\Omega,0)}{\partial \Omega^2}=-\frac{g^2}{4\pi^2}N_F F(\alpha)\frac{1}{\epsilon},
\end{equation}
where  function \(F(\alpha)\) is defined as 
\begin{equation}\label{eq:F}
    F(\alpha)=\frac{1}{(1-\alpha^2)^2}.
\end{equation}
With the above definition of the bosonic self-energy, the bosonic wave-function renormalization is determined by the renormalization condition
\begin{equation}
    Z_{\Phi}-\left.\frac{\partial\Sigma_B (i\Omega,0)}{\partial\Omega^2}\right|_{\Omega \to 0}=1,
\end{equation}
implying that the renormalization constant for the bosonic fields is 
\begin{equation}~\label{app:Zphi}
Z_{\Phi}=1+\left.\frac{\partial\Sigma_B (i\Omega,0)}{\partial\Omega^2}\right|_{\Omega \to 0}=1-\frac{g^2}{4 \pi^2}N_F F(\alpha)\frac{1}{\epsilon}.
\end{equation}

\subsection{Correction to Yukawa vertex}\label{App:Yukawa vertex}
The correction to the Yukawa vertex (Fig.~\ref{fig:diagramc}) reads 
\begin{equation}
\begin{aligned}
    \delta g=&g^2 \int[d\omega]\int [d^{D}\mathbf{q}]\left\{\left(P_{+}G_{F}(i\Omega_1+i\Omega_2 -i\omega,\mathbf{k+p-q})P_{+}G_F(i\Omega_1 -i\omega,\mathbf{k-q}) P_{-}\right)\right.\\
    &\left.+\left(P_{-}G_{F}(i\omega+i\Omega_1+i\omega_2,\mathbf{q+k+q})P_{+}G_F(i\omega+i\Omega_1,\mathbf{q+k}) P_{+}\right)\right.\\
    &+ \left(P_{+}G_{F}(i\Omega_1 -\Omega_2 -i\omega ,\mathbf{k-p-q})P_{-}G_F(i\Omega_1 -i\omega,\mathbf{k-q}) P_{-}\right)\\
    &\left.+\left(P_{-}G_{F}(i\Omega_1 -i\Omega_2 +i\omega ,\mathbf{k-p-q})P_{-}G_F(i\omega +i\Omega_1 ,\mathbf{q+k}) P_{+}\right)\right\}\frac{1}{\omega^2 + q^2}.
\end{aligned}
\end{equation}
Using the relation \(P_{\pm}P_{\pm}=0\), the form of the fermionic propagator in Eq.~\eqref{fermion-prop}, and that all the matrices either commute or anticommute with the matrices $P_\pm$, it can be readily shown that the correction of the Yukawa vertex identically vanishes,  $\delta g=0$, after taking all the external momenta and frequencies to be equal to zero. Therefore, the renormalization for the Yukawa coupling has the form 
\begin{equation}\label{gcondition}
    Z_{\phi}^{1/2}Z_{\Psi}g_0 \mu^{-\epsilon/2}=g,
\end{equation}
which, together with the form of the renormalization constants $Z_\Psi$ and $Z_\phi$, in Eqs.~\eqref{Zpsi} and~\eqref{app:Zphi}, respectively, implies the form of the RG flow equation of the Yukawa coupling in Eq.~\eqref{eq:beta-g2} of the main text.
\subsection{Correction to the $\phi^4$ vertex}\label{App4}

The first contribution to the \(\lambda\) coupling comes from the already know result form the \(\phi^4\) vertex (Fig.~\ref{fig:diagramd}), which reads
\begin{equation}
    \delta \lambda_{\lambda} = (N_B +4 )\frac{\lambda^2}{8\pi^2}\frac{1}{\epsilon},
\end{equation}
where, \(N_B\) is the number of complex components of the bosonic order parameter field. 

The correction to the \(\lambda\) coupling due to the Yukawa interaction (Fig.~\ref{fig:diagrame}), with
all external momenta and frequencies set to zero, reads
\begin{equation}
    \delta \lambda_{Y}=-g^4\int[d\omega]\int[d^D \mathbf{q}]\,\text{Tr}\left[P_{+}G_{F}(i\omega,\mathbf{q})P_{-}G_{F}(i\omega,\mathbf{q})P_{+}G_{F}(i\omega,\mathbf{q})P_{-}G_{F}(i\omega,\mathbf{q})\right].
\end{equation}
After taking the trace and integrate over the frequency and momentum, we obtain the divergent part of this correction 
\begin{equation}
    \delta \lambda_{Y}=-\frac{g^4}{ 2\pi^2} N_F F(\alpha)\frac{1}{\epsilon},
\end{equation}
where the function $F(\alpha)$ is defined in Eq.~\eqref{eq:F}. Finally, the renormalization condition of the \(\lambda\) reads
\begin{equation}\label{lambdaCondition}
    Z_{\phi}^2 \lambda_0 \mu^{-\epsilon}-(\delta\lambda_{Y}+\delta\lambda_\lambda)=\lambda,
\end{equation}
which ultimately yields the $\beta$ function for the $\lambda$ coupling in Eq.~\eqref{eq:beta-lambda}.

\bibliographystyle{JHEP}
\bibliography{Manuscript-SC-TDF}

\end{document}